\documentclass[amsmath,superscriptaddress,prd]{revtex4}

\usepackage{graphicx}

\begin{document}

\title{
{ 
 Optimal Renormalization-Group Improvement of the Perturbative Series for the 
$e^+e^-$-Annihilation Cross-Section
} }

\author{M.R.~Ahmady}
\affiliation{Department of Physics, Mount Allison University, Sackville, NB, E4L 1E6, Canada}
\affiliation{Dept.\ of Mathematics/Science, State Univ.\ of New York Institute of Technology, Utica, NY, 13504-3050, USA}
\affiliation{Department of Applied Mathematics,
The University of Western Ontario,
London, ON, N6A 5B7, Canada}
\author{F.A.~Chishtie}
\affiliation{Newman Laboratory of Nuclear Studies, Cornell University, 
Ithaca, NY, 14853, USA}
\author{V.~Elias}
\affiliation{Perimeter Institute for Theoretical Physics,
35 King Street North,
Waterloo, ON, N2J 2W9, Canada}
\affiliation{Department of Applied Mathematics,
The University of Western Ontario,
London, ON, N6A 5B7, Canada}
\author{A.H.~Fariborz} 
\affiliation{Dept.\ of Mathematics/Science, State Univ.\ of New York Institute of Technology, Utica, NY, 13504-3050, USA}
\author{D.G.C.~McKeon}
\affiliation{Department of Applied Mathematics,
The University of Western Ontario,
London, ON, N6A 5B7, Canada}
\affiliation{Dept.\ of Mathematics/Science, State Univ.\ of New York Institute of Technology, Utica, NY, 13504-3050, USA}
\author{T.N.~Sherry}
\affiliation{Department of Mathematical Physics, National University of Ireland, Galway, Ireland}
\affiliation{Dept.\ of Mathematics/Science, State Univ.\ of New York Institute of Technology, Utica, NY, 13504-3050, USA}
\author{A.\ Squires}
\affiliation{Department of Physics, Mount Allison University, Sackville, NB, E4L 1E6, Canada}
\affiliation{Dept.\ of Mathematics/Science, State Univ.\ of New York Institute of Technology, Utica, NY, 13504-3050, USA}
\author{T.G.~Steele}
\affiliation{Department of Applied Mathematics,
The University of Western Ontario,
London, ON, N6A 5B7, Canada}
\affiliation{Department of Physics \& Engineering Physics,
University of Saskatchewan,
Saskatoon, SK, S7N 5E2, Canada
}
\affiliation{Dept.\ of Mathematics/Science, State Univ.\ of New York Institute of Technology, Utica, NY, 13504-3050, USA}
 
\begin{abstract}
Using renormalization-group methods, we derive differential equations for the all-orders
summation of logarithmic corrections to the QCD series for $R(s) =\sigma(e^+e^- \to  {\rm hadrons})/\sigma(e^+e^-  
\to \mu^+\mu^-)$, as obtained from the imaginary part of the purely-perturbative vector-current correlation
function. We present explicit solutions for the summation of leading and up to three subsequent
subleading orders of logarithms. The summations accessible from the four-loop vector-correlator
not only lead to a substantial reduction in sensitivity to the renormalization scale, but necessarily
impose a common infrared bound on perturbative approximations to  $R(s)$, regardless of the infrared behaviour of the true 
QCD couplant.
\end{abstract}

\maketitle

For center-of-mass squared-energy $s$, QCD corrections to 
$R(s) \equiv     
\sigma(e^+ e^-\to {\rm hadrons})/ \sigma(e^+ e^-\to \mu^+ \mu^-)$ are
scaled by a perturbative QCD series ($S$):
\begin{equation}\label{R_def}
R(s) = 3 \sum_f Q_f^2 \,S\left[ \frac{\alpha_s (\mu)}{\pi} , \log \left(\frac{\mu^2}{s}\right) \right].
\end{equation}
This series is extracted from the imaginary part of the $\overline{{\rm MS}}$ vector-current 
correlation function \cite{sgg,ces},
\begin{equation}\label{S_def}
S[x,L] = 1 + \sum_{n=1}^\infty x^n \sum_{m=0}^{n-1} T_{n,m} L^m ,
\end{equation}
with coefficients $T_{n,m}$ tabulated in Table \ref{Table_I} for 3--5 active flavors, as 
appropriate for the choice of the center-of-mass squared-energy $s$.  Each 
order of this series depends upon the $\overline{{\rm MS}}$ renormalization scale 
parameter $\mu$, both through the couplant
\begin{equation}\label{x_def}
x(\mu) \equiv \alpha_s (\mu)/\pi
\end{equation}
and through powers of the logarithm
\begin{equation}\label{L_def}
L(\mu) \equiv \log (\mu^2 / s).
\end{equation}
Nevertheless, the all-orders series $S$ must ultimately be independent of 
renormalization scale. $R(s)$ is a measurable physical quantity necessarily 
independent of $\mu$, the artificial scale entering QCD calculations as a 
by-product of the regulation of Feynman-diagrammatic infinities.  Hence, 
\begin{equation}\label{RG_eq}
0 = \mu^2 \frac{\mathrm{d}S[x(\mu), L(\mu)]}{\mathrm{d}\mu^2} = \left( \frac{\partial}{\partial L} 
+ \beta (x) \frac{\partial}{\partial x} \right) S[x,L].
\end{equation}
The above renormalization group equation (RGE) is simply a chain-rule 
relation in which
\begin{equation}\label{beta_def}
\beta(x) \equiv \mu^2 \frac{\mathrm{d}x(\mu)}{\mathrm{d}\mu^2} = -x^2 \sum_{k=0}^\infty \beta_k x^k ,
\end{equation}
where known \cite{rvl} $\overline{{\rm MS}}$ $\beta$-function coefficients $\beta_k$ are also 
tabulated in Table \ref{Table_I}.  Thus, the RGE (\ref{RG_eq}) is generally employed to provide scale dependence to the
couplant $x$.  

\begin{table}
\centering
\begin{tabular}{||c|c|c|c||}
\hline \hline
 & $n_f=3$ & $n_f=4$   & $n_f=5$ \\ \hline \hline
$T_{1,0}$ & $1$ & $1$  & $1$ \\ \hline
$T_{2,0}$ & $1.63982$  & $1.52453$ &  $1.40924$ \\ \hline
$T_{2,1}$ & $9/4$      & $25/12$   &  $23/12$ \\ \hline
$T_{3,0}$ & $-10.2839$ & $-11.6856$   & $-12.8046$ \\ \hline
$T_{3,1}$ & $11.3792$  & $9.56054$    & $7.81875$ \\ \hline
$T_{3,2}$ & $81/16$    & $625/144$    & $529/144$ \\ \hline
$\beta_0$ & $9/4$      & $25/12$      &  $23/12$ \\ \hline
$\beta_1$ & $4$        & $77/24$      &  $29/12$ \\ \hline
$\beta_2$ & $3863/384$ & $21943/3456$ &  $9769/3456$ \\ \hline
$\beta_3$ & $47.2280$  & $31.3874$    &  $18.8522$\\  \hline \hline
\end{tabular}
\caption{Coefficients for the imaginary part of the four-loop-order vector-current
correlation function, as well as coefficients for the four-loop order $\overline{{\rm MS}}$ $\beta$-function, are
listed for three, four and five quark flavors.}
\label{Table_I}
\end{table} 

``Optimal'' renormalization-group (RG) improvement is the inclusion of every term in a perturbative series of the form (\ref{S_def}) 
that can be extracted by RG-methods from a perturbative computation to a given order \cite{utica}.  For example, a 
next-to-next-to-leading (NNL) order 
perturbative calculation determines only the coefficients $T_{1,0}$, $T_{2,0}$, and $T_{2,1}$  of the series (\ref{S_def}).  
However, the RGE (\ref{RG_eq}) can be utilized to determine all coefficients $T_{k+1,k}$ and $T_{k+2,k}$ within the 
series (\ref{S_def}).  The contributions of this infinite set of coefficients may then be summed analytically, as described below, 
thereby providing an ``optimal RG-improvement'' of the NNL expression.

The series $S[x,L]$, as defined in Eq.\ (\ref{S_def}), may be rearranged 
in the following form:
\begin{equation}\label{S_series}
S[x,L] = 1 + \sum_{n=1}^\infty x^n S_n (xL),
\end{equation}
where
\begin{equation}\label{S_k_def}
S_k (u) \equiv \sum_{n=k}^\infty T_{n, n-k} u^{n-k}.
\end{equation}
Given knowledge of the $k^{th}$-order series coefficient $T_{k,0} = S_k (0)$, one 
can obtain $S_k (x(\mu)L(\mu))$ explicitly, thereby summing over the entire set 
of $k^{th}$-order subleading logarithms contributing to the series (\ref{S_series}). If 
we substitute the  $\beta$-function series (\ref{beta_def}) into the RGE (\ref{RG_eq}), we find that 
the aggregate coefficient of $x^n L^{n-p}$ vanishes $(n \geq p)$ provided the 
following recursion relation is upheld:
\begin{equation}\label{recursion}
0 = (n-p+1) T_{n, n-p+1} - \sum_{\ell = 0} ^{p-2} (n - \ell - 1) \beta_\ell T_{n - \ell - 1, n - p}.
\end{equation}
For example, if $p = 2$, this recursion relation $[T_{n,n-1} = \beta_0 T_{n-1,n-2}]$ 
relates all leading-logarithm coefficients $T_{n,n-1}$ within the series (\ref{S_def}) 
to the known coefficient $T_{1,0} = 1$, thereby enabling one to sum all 
orders of the leading-logarithm contributions
\begin{equation}
x S_1 (xL) = x \sum_{n=1}^\infty T_{n, n-1} (xL)^{n-1} = \frac{x}{1 - \beta_0 xL}
\end{equation}
to the series $S[x,L]$.

More generally, the recursion relation (\ref{recursion}) may be utilized to obtain 
a succession of first-order inhomogeneous linear differential equations 
for the functions $S_k (u)$ within Eq.\ (\ref{S_series}). If one multiplies Eq.\ (\ref{recursion}) by $u^{n-p}$ and then sums 
over $n$ from $n = p$ to $\infty$, one finds from the definition (\ref{S_k_def}) of $S_k (u)$ that
\begin{equation}
0  =  \frac{\mathrm{d} S_{p-1}}{\mathrm{d}u} - u \sum_{\ell = 0}^{p-2} \beta_\ell \frac{\mathrm{d} S_{p-\ell - 1}}{\mathrm{d}u}
 -  \sum_{\ell = 0}^{p-2} (p - \ell - 1) \beta_\ell S_{p - \ell - 1},
\end{equation}
which can be trivially rearranged $(k = p - 1)$ into a set of first-order 
linear differential equations
\begin{equation}\label{S_k_de}
\frac{\mathrm{d}S_k}{\mathrm{d}u} - \frac{k\beta_0}{(1-\beta_0 u)} S_k = \frac{1}{(1-\beta_0 u)} \sum_{\ell = 1}^{k-1} \beta_\ell \left( u \frac{\mathrm{d}}{\mathrm{d}u} + k - \ell \right) S_{k - \ell},
\end{equation}
with initial conditions $S_k (0) = T_{k,0}$. 

Noting that $S_0 (u) = 1$ and that $T_{1,0} = 1$ regardless of the number of active flavors, we 
find the first four solutions of (\ref{S_k_de}) to be
\begin{gather}\label{S1}
S_1 (xL) = \frac{1}{(1 - \beta_0 xL)},
\\
\label{S2}
S_2 (xL) = \frac{T_{2,0} - \frac{\beta_1}{\beta_0} \log 
\left( 1 - \beta_0 xL \right)}{\left( 1-\beta_0 xL \right)^2},
\\
\begin{split}
S_3 (xL)  = & \left( \frac{\beta_1^2}{\beta_0^2} - \frac{\beta_2}{\beta_0} \right) / \left( 1 - \beta_0 xL \right)^2 
\\
& +  \frac{ \left\{T_{3,0} - \left( \frac{\beta_1^2}{\beta_0^2} - \frac{\beta_2}{\beta_0} \right)  
- \frac{\beta_1}{\beta_0} \left( 2 T_{2,0} + \frac{\beta_1}{\beta_0} \right) \log (1 - \beta_0 xL)
+ \frac{\beta_1^2}{\beta_0^2} \log^2 (1 - \beta_0 xL) \right\} } {\left( 1 - \beta_0 xL \right)^3} 
\end{split}
\\
\begin{split}\label{S4}
S_4 (xL)  = & -\frac{1}{2} \left[ \frac{\beta_1}{\beta_0} \left( \frac{\beta_1^2}{\beta_0^2} - 2 \frac{\beta_2}{\beta_0} \right) + \frac{\beta_3}{\beta_0} \right] (1 - \beta_0 xL )^{-2} 
 +  \left( 2 T_{2,0} + \frac{\beta_1}{\beta_0} \right) \left( \frac{\beta_1^2}{\beta_0^2} - \frac{\beta_2}{\beta_0} \right) (1 - \beta_0 xL)^{-3} 
\\
& +  2 \frac{\beta_1}{\beta_0} \left( \frac{\beta_2}{\beta_0} - \frac{\beta_1^2}{\beta_0^2} \right) (1 - \beta_0 xL)^{-3} \log (1 - \beta_0 xL) 
\\
& +  \left[ T_{4,0} + \frac{\beta_3}{2\beta_0} - \frac{1}{2} \frac{\beta_1^3}{\beta_0^3} - 2T_{2,0} \left( \frac{\beta_1^2}{\beta_0^2} - \frac{\beta_2}{\beta_0} \right) \right] (1 - \beta_0 xL)^{-4}\\
& +  \frac{\beta_1}{\beta_0}\left[ 2\frac{\beta_1^2}{\beta_0^2} - 3 \frac{\beta_2}{\beta_0} - 2T_{2,0} \frac{\beta_1}{\beta_0} - 3T_{3,0} \right] (1 - \beta_0 xL)^{-4} \log (1 - \beta_0 xL)
\\
& +  \frac{\beta_1}{\beta_0} \left[ \frac{5\beta_1^2}{2\beta_0^2} + 3T_{2,0} \frac{\beta_1}{\beta_0}\right] (1 - \beta_0 xL)^{-4} \log^2 (1 - \beta_0 xL)
 -  \frac{\beta_1^3}{\beta_0^3} (1 - \beta_0 xL)^{-4} \log^3 (1 - \beta_0 xL). 
\end{split}
\end{gather}

To explore the near-infrared regime of perturbative QCD, we specialize 
to the case of three active flavors. Using Table \ref{Table_I} $n_f = 3$ values within 
Eqs.\ (\ref{S1})--(\ref{S4}), we find that the version of the series (\ref{S_series}) which 
incorporates full summation of leading and two subsequent subleading 
orders of logarithms is given by
\begin{equation}\label{S_Sigma}
S^{(\Sigma)} \left[ x(\mu), L(\mu) \right] = 1 + x S_1 (xL) + x^2 S_2 (xL) + x^3 S_3 (xL),
\end{equation}
where $(u = xL)$
\begin{gather}\label{S1_3f}
S_1 (u) = 1 / (1 - 9u / 4),
\\
S_2 (u) = \frac{1.63982 - \frac{16}{9} \log (1 - 9u / 4)}{(1 - 9u / 4)^2},
\\
\label{S3_3f}
S_3 (u)  =  - \frac{1.31057}{(1 - \frac{9}{4} u)^2}
 +  \frac{\left\{ -8.97333 - 8.99096 \log \left( 1 - \frac{9u}{4} \right) + 3.16049 \log^2 \left( 1 - \frac{9u}{4} \right) \right\}}{(1 - \frac{9u}{4})^3}.
\end{gather}
Moreover, we note from Eq.\ (\ref{S4}) that 
\begin{equation}
\begin{split}
 S_4 (xL)  =&  -\frac{5.35589}{ \left(1 - \frac{9}{4}xL\right)^{2}} 
 +   \frac{\left[ -6.62811 + 4.65981 \log\left(1 - \frac{9}{4}xL\right) \right]} 
{\left(1 - \frac{9}{4}xL\right)^{3}} 
\\
& +  \frac{\left[ T_{4,0} + 11.9840 + 31.8738 \log\left(1 - \frac{9}{4}xL\right) 
 +   29.5946 \log^2\left(1 - \frac{9}{4}xL\right) 
- 5.61866 \log^3\left(1 - \frac{9}{4}xL\right) \right]}{\left(1 - \frac{9}{4}xL\right)^{4}} \quad,
\end{split}
\label{S4_3f}
\end{equation}
thereby providing for inclusion of the $x^4 S_4 (xL)$ contribution to the series (\ref{S_series}) 
 for three active flavors.
The series coefficient $T_{4,0}$ appearing in Eq.\ (\ref{S4_3f}) has not yet been 
calculated perturbatively, which is why we have not included the 
$x^4 S_4 (xL)$ contribution to $S[x,L]$ in Eq.\ (\ref{S_Sigma}).
(An asymptotic Pad\'e approximant
estimate $T_{4,0} \simeq 1.90$ for the $n_f = 3$ case is presented in ref. \cite{ces}.)

\begin{figure}[hbt]
\centering
\includegraphics[scale=0.4]{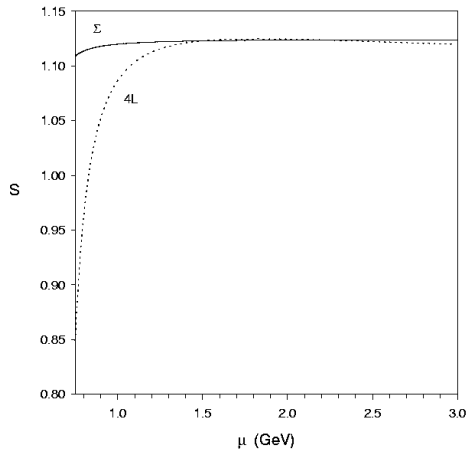}
\caption{The renormalization-scale ($\mu$) dependence of the summation-of-logarithms
($\Sigma$) series (\protect\ref{S_Sigma}) is compared to that of the series (\protect\ref{S_4l}), which is truncated
after four-loop ($4L$) contributions to the vector-current correlation function.
For both series the physical momentum scale $s$ is fixed at $1.5 \; {\rm GeV}^2$.  The
evolution of the couplant $x(\mu)$ for both series is referenced to the initial value
$\alpha_s (m_{\tau}) = 0.33$.}
\label{fig1}
\end{figure}

To examine whether the summation of leading and subsequent subleading 
logarithm factors decreases dependence on the unphysical 
renormalization-scale parameter $\mu$, we compare the $\mu$-dependence of 
Eq.\ (\ref{S_Sigma}) for a fixed value of $s$ to that of the $n_f =3$ version of the 
series (\ref{S_def}) truncated after four-loop-order ($4\ell$) contributions to the vector-current
correlation function:
\begin{equation}\label{S_4l}
S^{(4\ell)} \left[ x(\mu), L(\mu) \right]  =  1 + x + (1.63982 + 9L / 4) x^2 
 +  (-10.2839 + 11.3792L + 81L^2 / 16) x^3.
\end{equation}
Such $\mu$-dependence enters Eqs.\ (\ref{S_Sigma}) and (\ref{S_4l}) both through $L = \log(\mu^2/s)$ and 
through $x = x(\mu)$, which is assumed to evolve via Eq.\ (\ref{beta_def}) (with $n_f = 3$ 
choices for $\beta_{0-3}$) from an initial value 
choice $x(m_{\tau}) = \alpha_s(m_{\tau})/\pi = 0.33/\pi$ \cite{aleph, ptg}.  Figure \ref{fig1} displays 
a comparison of the $\mu$-dependence of Eqs.\ (\ref{S_Sigma}) and (\ref{S_4l}) at 
fixed $s = 1.5 \; {\rm GeV}^2$ .  Although both expressions exhibit little 
variation with $\mu$ over the $1.3 \; {\rm GeV} \leq \mu \leq 3 \; {\rm GeV}$ range, we see that Eq.\ 
(\ref{S_Sigma}) exhibits much less variation with $\mu$ in the near-infrared regime 
below $1.3 \; {\rm GeV}$. These results clearly indicate that 
renormalization-scale-invariance is more effectively upheld via 
the summations of leading and subsequent subleading orders of 
logarithms that occur within Eq.\ (\ref{S_Sigma}).  We emphasize that Eqs.\ (\ref{S_Sigma}) 
and (\ref{S_4l}) both follow from ``RG-improvement'' of the same calculational 
information [the coefficients $T_{1,0}, \; T_{2,0},$ and $T_{3,0}$]; simply 
put, such RG-improvement is more effectively implemented in Eq.\ (\ref{S_Sigma}) 
than in Eq.\ (\ref{S_4l}).

\begin{figure}[hbt]
\centering
\includegraphics[scale=0.4]{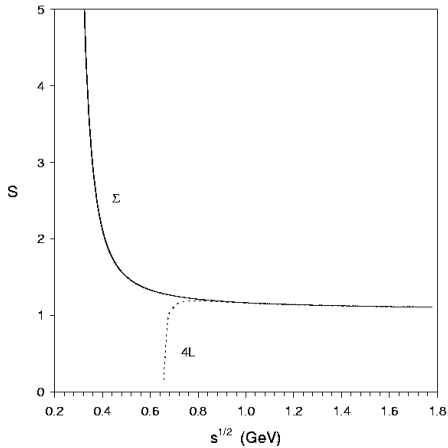}
\caption{The center-of-mass squared-energy ($s$) dependence of the four-loop ($4L$-) truncated 
series (\protect\ref{S_4l_std}) is compared to that of the summation-of-logarithms ($\Sigma$) series (\protect\ref{S_Sigma_tau}),
as described in the text.  In both series, $\alpha_s (m_{\tau})$ is taken to be $0.33$ so that the
series equilibrate at $\sqrt{s} = m_{\tau}$.}
\label{fig2}
\end{figure}

The usual prescription for obtaining the purely-perturbative (non-power-law) QCD 
contributions to $R(s)$ at four-loop order \cite{ptg, fjy} is to set 
$\mu = \sqrt{s}$ within Eq.\ (\ref{S_4l}), and then to substitute the resulting series,
\begin{equation}\label{S_4l_std}
S^{(4\ell)} \left[ x(\sqrt{s}), L(\sqrt{s}) \right] = 1 + x (\sqrt{s}) + 1.63982 \; x^2 (\sqrt{s}) - 10.2839 \; x^3 (\sqrt{s}),
\end{equation}
into Eq.\ (\ref{R_def}). [Note from Eq.\ (\ref{L_def}) that $L(\sqrt{s}) = 0$.] This prescription 
follows from the presumed renormalization-scale invariance of the 
truncated series (\ref{S_4l}), thereby leading to an expression that depends 
only on the physical scale $s$ for the electron-positron annihilation 
process.
Note that all $s$-dependence of Eq.\ (\ref{S_Sigma}) resides entirely in the 
variable $L$. Let us first fix $\mu = m_\tau$  within Eq.\ (\ref{S_Sigma}) so as to incorporate 
the benchmark couplant value   $x(m_\tau) = 0.33/\pi$  everywhere $x$ appears in 
Eqs.\ (\ref{S1_3f})--(\ref{S3_3f}). With this choice, 
the following summation-of-logarithms series 
can be substituted into Eq.\ (\ref{R_def}) to obtain $R(s)$:
\begin{equation}\label{S_Sigma_tau}
\begin{split}
S^{(\Sigma)} \left[x (m_\tau), L(m_\tau) \right]  = & 1 + \frac{0.33}{\pi} 
S_1 \left( \frac{0.33}{\pi} \log \left( \frac{m_{\tau}^2}{s} \right) \right) 
\\
& +  \left( \frac{0.33}{\pi} \right)^2 S_2 \left( \frac{0.33}{\pi} \log \left(\frac{m_\tau^2}{s} \right) \right)  
 +  \left( \frac{0.33}{\pi}\right)^3 S_3 \left( \frac{0.33}{\pi} \log \left( \frac{m_\tau^2}{s} \right) \right) .
\end{split}
\end{equation}
In Figure \ref{fig2} we compare the $s$-dependence of this series to that of 
Eq.\ (\ref{S_4l_std}), for which all $s$ dependence resides in the evolution 
of $x(\sqrt{s})$.  To make this comparison, such evolution is 
anchored to the initial value $x(m_{\tau}) = 0.33/\pi$ via the differential 
equation (\ref{beta_def}) with $n_f = 3$ values for $\beta_{0-3}$ [Table \ref{Table_I}]. This initial 
value ensures that the series (\ref{S_4l_std}) and (\ref{S_Sigma_tau}) coincide when $\sqrt{s} = m_{\tau}$. 
Figure \ref{fig2} shows that both series continue to coincide over the range 
$750 \; {\rm MeV} \leq \sqrt{s} \leq m_{\tau}$. For values of $\sqrt{s}$ less than 750 {\rm MeV}, however, 
the truncated series (\ref{S_4l_std}) drops off quite suddenly at  $\sqrt{s} \simeq 650 \; {\rm MeV}$, a consequence of the large negative coefficient of $x^3\left(\sqrt{s}\right)$,
whereas the series (\ref{S_Sigma_tau}) [as obtained from the full summation-of-logarithms 
series (\ref{S_Sigma})] continues to probe the infrared domain of $R(s)$ even for 
values of  $\sqrt{s} \simeq 400 \; {\rm MeV}$. In short, the summation of all leading and 
subsequent two subleading logarithms within the perturbative 
series (\ref{S_def}) serves to extend the domain of the $R(s)$ series further into the 
infrared. This property, as well as the reduced renormalization-scale 
dependence evident in Figure \ref{fig1}, suggests that such summation is 
particularly appropriate for the near-infrared region characterizing 
sum-rule applications of purely-perturbative QCD corrections to
current-correlation functions.

It is also evident from Figure \ref{fig2} that the domain of the 
summation-of-logarithms series (\ref{S_Sigma_tau}) manifests a singularity  
below  $\sqrt{s} = 400 \; {\rm MeV}$, despite the fact that the $s$-dependence of the 
series (\ref{S_Sigma_tau}) is decoupled entirely from any infrared behavior of 
the couplant $x$, which is held constant at $x(m_{\tau})$. To understand this restriction 
on the domain of $R(s)$, we first note that each summation 
(\ref{S1})--(\ref{S4}) becomes singular when $1 - \beta_0 xL \rightarrow 0$. 
Such resummation singularities have also been observed to occur in 
completely different contexts, 
including deep inelastic structure functions \cite{vogt}.

\begin{figure}[hbt]
\centering
\includegraphics[scale=0.4]{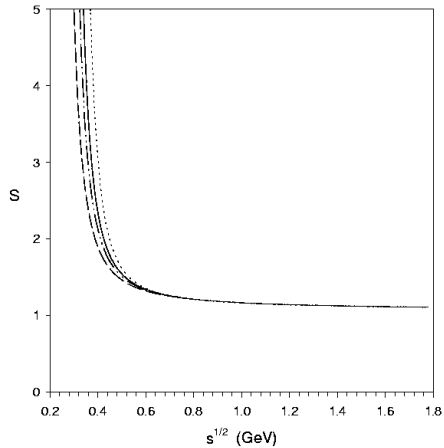}
\caption{
The $s$ dependence of the four-loop truncated series  (\protect\ref{S_Sigma})
for equally-spaced values of the renormalization scale in the range 
$0.6 m_\tau\le\mu\le 1.6 m_\tau$.  The couplant $x(\mu)$ is obtained from three-flavour four-loop
evolution via (\protect\ref{beta_def})   
from the initial condition $x\left(m_\tau\right)=0.33/\pi$ using  
 Table \protect\ref{Table_I} values for the  $\beta$ function coefficients
$\beta_0$--$\beta_3$.
}
\label{fig3}
\end{figure}

The singularity  property of eqs.\  (\ref{S1})--(\ref{S4}) 
is  upheld for {\em all} summations $S_k(xL)$. The solution to the differential 
equation (\ref{S_k_de}) is necessarily of the form
\begin{equation}\label{S_solution}
S_k (xL) = \frac{T_{k,0}}{(1 - \beta_0 xL)^k} + \left( \mbox {Particular solution depending on} \left\{ S_{k-1}, S_{k-2}, ... , S_1 \right\} \right).
\end{equation}
Since the coefficients $T_{k,0}$ are results of k$^{th}$-order Feynman diagram 
calculations, the k$^{th}$-order pole in (\ref{S_solution}) at $1 - \beta_0 xL = 0$ is 
genuine and will not be canceled by particular-solution contributions 
that are sensitive to at most $(k-1)^{th}$-order Feynman-diagrammatic 
coefficients $\{ T_{1,0},~ T_{2,0}, \ldots,~ T_{k-1, 0} \}$. 
For a given choice of renormalization scale $\mu$, this 
singularity implies  [via Eqs.\ (\ref{x_def}) and (\ref{L_def})] that each summation $S_k (xL)$ 
within the full series (\ref{S_series}) becomes singular for a 
sufficiently small value of $s$:
\begin{equation}\label{s_min}
1 - \beta_0 \frac{\alpha_s (\mu)}{\pi} \log \left( \frac{\mu^2}{s_{min}} \right) = 0 \quad
\rightarrow s_{min} = \mu^2 \exp \left( \frac{-\pi}{\beta_0 \alpha_s (\mu)} \right).
\end{equation}
For example, if the renormalization scale $\mu$ is chosen (as in Figure \ref{fig2}) 
to be $m_{\tau}$, a choice for which $\alpha_s (m_{\tau}) (= 0.33 \pm 0.02$ \cite{ptg}) is 
phenomenologically accessible, then each term in the series (\ref{S_series}) is 
seen to become progressively more singular as $s$ approaches 
$m_{\tau}^2 \exp[-4\pi /(9 \times 0.33)] = (215 \; {\rm MeV})^2$ from above. 
Furthermore, Figure \ref{fig3} shows that the low-energy behaviour of the resummed expression
(\ref{S_Sigma}) is only weakly dependent on $\mu$ in the region $\sqrt{s}>600\,{\rm MeV}$ for
choices of $\mu$ between 
$0.6m_\tau$ and $1.6m_\tau$, 
with the appearance of singular points  corresponding to 
$195\,{\rm MeV}<s_{min}<250\,{\rm MeV}$ from (\ref{s_min}).

It is to be emphasized that this infrared boundary on the physical 
scale $s$ entering term-by-term within the  series (\ref{S_series}) is {\em not} a manifestation 
of any infrared boundary \cite{cem} on the evolution of the QCD couplant $x(\mu)$.
Even if the higher order contributions to the $\beta$-function (\ref{beta_def}) were somehow to conspire 
to allow the couplant to be well-behaved in the infrared region 
[{\it e.g.,} to have infrared-stable fixed point behavior], the restriction (\ref{s_min}) 
would still apply upon making a specific choice of the 
renormalization-scale parameter $\mu$ and its corresponding value of $\alpha_s (\mu)$. Curiously, though, this infrared 
restriction on $s$ can be easily shown to coincide with the 
``infrared-slavery'' Landau singularity $\Lambda$ associated with 
naive evolution of the QCD couplant via a one-loop  $\beta$-function.  The one 
loop version of Eq.\ (\ref{beta_def}),
\begin{equation}
\mu^2 \frac{\mathrm{d}x}{\mathrm{d}\mu^2} = -\beta_0 x^2, \; \; x=\frac{\alpha_s}{\pi},
\end{equation} 
is satisfied by the relation
\begin{equation}
\frac{\alpha_s}{\pi} = \frac{1}{\beta_0 \log (\mu^2 / \Lambda^2)}
\end{equation}
which is equivalent to Eq.\ (\ref{s_min}) provided  $s_{min}$ is identified with $\Lambda^2$. Indeed, in a one-loop 
world [$x^{1\ell}(\mu)=1/\beta_0\log\left(\mu^2/\Lambda^2\right)$] where $\Lambda$ serves as a {\em universal} infrared boundary, 
the  one-loop analogues of the
 summation-of-logarithms series (\ref{S_Sigma}) and the truncated series (\ref{S_4l_std}) are necessarily equivalent:
\begin{equation}
\begin{split}
S_{1\ell}^{(\Sigma)}&=1+x^{(1\ell)}(\mu)
S_1\left[x^{(1\ell)}(\mu)\log\left(\mu^2/s\right)\right]
=1+\frac{1}{\beta_0\log\left(\frac{\mu^2}{\Lambda^2}\right)}\left\{
\frac{1}{1-\beta_0\left[\frac{1}{\beta_0\log\left(\frac{\mu^2}{\Lambda^2}\right)}\right]\log\left(\frac{\mu^2}{s}\right) } \right\}
\\
&=1+\frac{1}{\beta_0\log\left(\frac{s}{\Lambda^2}\right)}=1+x^{(1\ell)}\left(\sqrt{s}\right)
\end{split}
\end{equation}
We find it remarkable that $\Lambda$, the one-loop couplant's Landau pole, persists as an infrared boundary 
on the domain of each summation contributing 
to Eq.\ (\ref{S_series}), 
the summation-of-logarithms formulation of the perturbative series 
within $R(s)$.  Consequently, $s=\Lambda^2$ serves as an infrared boundary for any approximation to $R(s)$
involving the truncation of the series (\ref{S_series}), such as the expression (\ref{S_Sigma}) obtained via optimal 
RG-improvement of the four-loop vector-current correlation function's imaginary part.

\begin{acknowledgments}
 We are grateful to V.A.\ Miransky for useful discussions, and to the Natural Sciences \& Engineering 
Research Council of Canada (NSERC) for financial support.  We also wish to thank SUNY Institute of Technology for hospitality in the
preparation of this research.  The work of AHF has been supported in part by the 2002 State of New York/UUP Professional Development 
Committee, and the 2002 Summer Grant from the School of Arts and Sciences, SUNY Institute of Technology. 
The work of MRA has been  supported in part by an internal summer research grant from Mount Allison University. AS wishes to acknowledge 
support from an NSERC Summer Undergraduate Research Award.
\end{acknowledgments}






\end{document}